\documentclass[journal=jacsat,manuscript=article]{achemso}

\usepackage[version=3]{mhchem} % Formula subscripts using \ce{}

\newcommand{\rfig}[1]{Fig.~\ref{#1}}
\newcommand{\req}[1]{Eq.~(\ref{#1})}

\title{Facile and fast growth of high mobility nanoribbons of ZrTe$_5$}

\author{Jingyue Wang}
\author{Jingjing Niu}
\author{Xinqi Li}
\author{Xiumei Ma}
\affiliation{State Key Laboratory for Artificial Microstructure and Mesoscopic Physics Peking University, Beijing 100871, China}
\author{Yuan Yao}
\affiliation{Institute of Physics, Beijing 100190, China}
\author{Dapeng Yu}
\affiliation{State Key Laboratory for Artificial Microstructure and Mesoscopic Physics Peking University, Beijing 100871, China}
\alsoaffiliation{Collaborative Innovation Center of Quantum Matter, Beijing 100871, China}
\alsoaffiliation{Department of Physics, South University of Science and Technology of China, Shenzhen 518055, China}
\author{Xiaosong Wu}
\email{xswu@pku.edu.cn}
\affiliation{State Key Laboratory for Artificial Microstructure and Mesoscopic Physics Peking University, Beijing 100871, China}
\alsoaffiliation{Collaborative Innovation Center of Quantum Matter, Beijing 100871, China}
\alsoaffiliation{Department of Physics, South University of Science and Technology of China, Shenzhen 518055, China}

\begin{document}

\begin{abstract}
 Recently, ZrTe$_5$ has received a lot of attention as it exhibits various topological phases, such as weak and strong topological insulators, a Dirac semimetal, and a quantum spin Hall insulator in the monolayer limit. While most of studies have been focused on the three-dimensional bulk material, it is highly desired to obtain nanostructured materials due to their advantages in device applications. We report the synthesis and characterizations of ZrTe$_5$ nanoribbons. Via a silicon-assisted chemical vapor transport method, long nanoribbons with thickness as thin as 20 nm can be grown. The growth rate is over an order of magnitude faster than the previous method for growth of bulk crystals. Moreover, transport studies show that nanoribbons are of low unintentional doping and high carrier mobility, over 30,000 cm$^2$/Vs, which enable reliable determination of the Berry phase of $\pi$ in the $ac$ plane from quantum oscillations. Our method holds great potential in growth of high quality ultra-thin nanostructures of ZrTe$_5$.
\end{abstract}

\maketitle

The discovery of topological insulators(TI) has brought a new insight into the classification of solid state materials and attracted enormous interest in the past few years\cite{Hasan2010,Qi2011}. Soon, the topological concept has been extended to superconductors and metals, \textit{i.e.}, three-dimensional(3D) topological Dirac semimetals and Weyl semimetals\cite{Kitaev2009,Roy2008,Schnyder2008,Qi2009,Hasan2010,Qi2011,Wan2011,Burkov2011}. Among many topological materials, ZrTe$_5$ is unique as it exhibits various topological phases, such as weak and strong topological insulators, a Dirac semimetal, and a quantum spin Hall insulator in the monolayer limit\cite{Zheng2016,Yu2016,Zhou2015,Chen2015,Li2014,Yuan2015,Wu2016,Pariari2017,Chen2015a,Manzoni2016,Li2016}. These phases sensitively depend on the lattice constants\cite{Weng2014}. Change of the band structure with temperature, thickness and pressure has been experimentally observed\cite{Niu2017,Lu2017,Zhang2017,Zhang2017a}.

Most of studies on ZrTe$_5$ have been carried out for 3D bulk\cite{Chi2017,Pariari2017,Liu2016,Yuan2015,Li2014}. On the other hand, nanowires/nanoribbons of topological materials can be very useful, as they not only enhance the contribution from surface states due to a large surface-to-volume ratio, but also give rise to new properties\cite{Imura2012,Qian2014,Zhou2014}. For instance, studies on nanostructured topological insulators have revealed interesting phase and spin related transport\cite{Zhang2010,Peng2010,Hong2014}. The marriage between InSb nanowires and superconductors has given birth to the experimental realization of Majarona zero modes\cite{Mourik2012}. It is therefore highly desired to develop methods for growth of high quality nanostructured ZrTe$_5$ so that new properties can be further introduced. However, such a study has so far not been reported.

Conventionally, ZrTe$_5$ were synthesized by a chemical vapor transport method. The growth is time-consuming, in the order of weeks, which hinders efforts on improvement of the crystal quality\cite{Sodeck1979,Levy1983,Li2014,Wu2016}. In this work, we employ a silicon-assisted chemical vapor transport method to grow both ZrTe$_5$ nanoribbons and bulk crystals. Growth time of millimeter size single crystals is reduced to less than 90 minutes, representing substantial improvement of the growth rate. Quantum transport measurements show that the as-grown nanoribbons are of high mobility and low unintentional doping. Well-resolved quantum oscillations confirm a Dirac band with a Berry phase $\pi$ in the $ac$ plane. Our method holds great potential in growth of ultra-thin nanostructures and single crystals of ZrTe$_5$.

\begin{figure}[htbp]
\includegraphics[width=1\columnwidth]{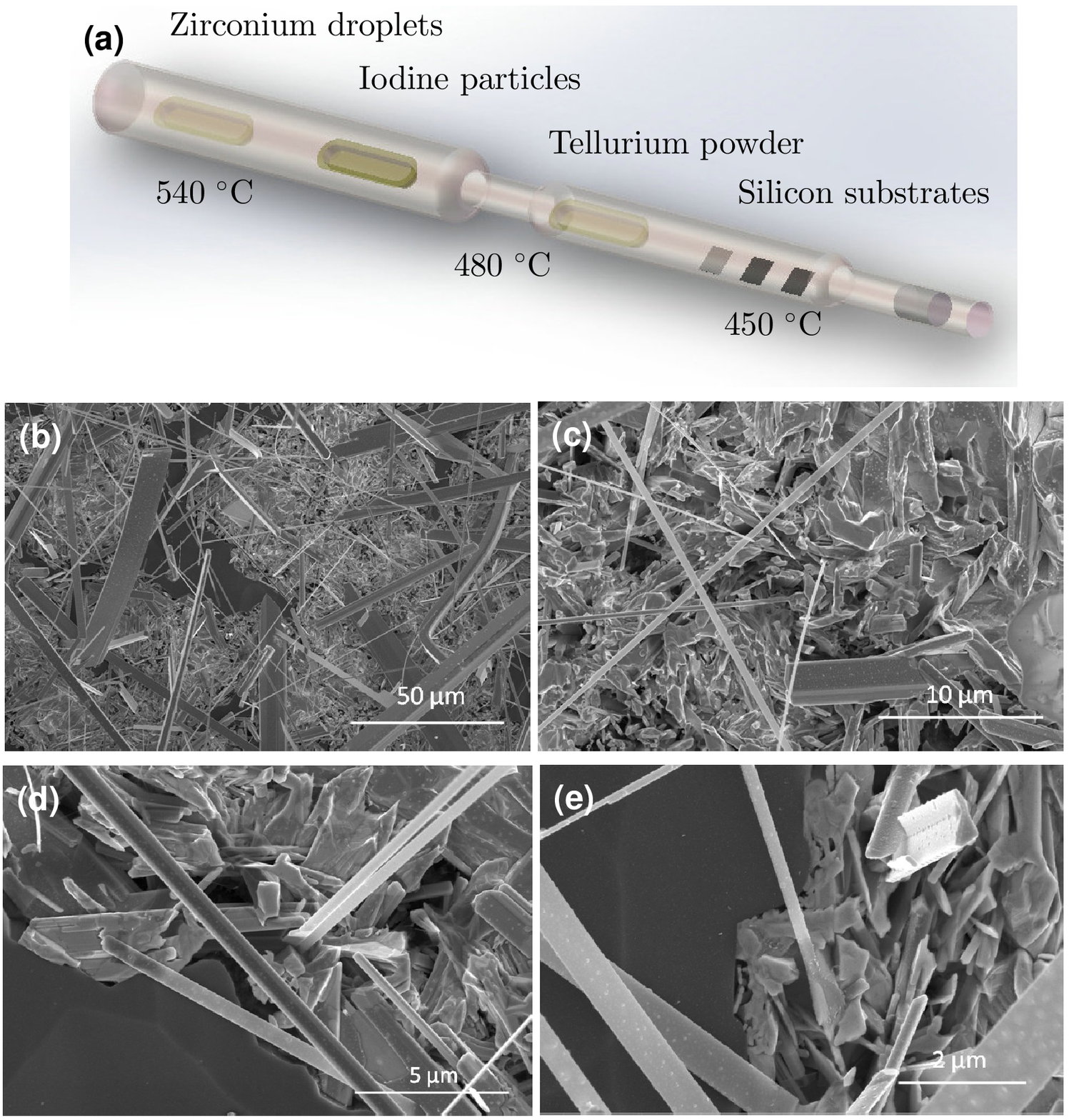}
\caption{Growth of ZrTe$_5$ nanostructures. (a)A schematic drawing of the growth setup for the iodine vapor transport method. A two-bulb ampoule is used. Zirconium shots and iodine particles are placed in one bulb, which is held at a higher temperature, 540 $^\circ$C. Tellurium powder and silicon substrates are placed in the other bulb, which is held at a lower temperature, 480 $^\circ$C and 450 $^\circ$C, respectively. (b)(c)(d)(e)A series of SEM images of grown nanostructures with increasing magnifications. ZrTe$_5$ nanoribbons are grown on a mattress of ZrTe$_3$ and Te crystals. A recession of the silicon substrate due to iodine etching is marked by arrows in (d) and (e).}
\label{fig.sem}
\end{figure}

Growth was carried out in a horizontal three-zone tube furnace. Source materials are zirconium and tellurium elements. Since zirconium powder is difficult to handle as it is easily oxidized and flammable, shots of 0.6 g each on average were used instead. Considering much less surface area of shots than powder, the amount of zirconium is significantly more than the stoichiometric ratio, e.g. 5 g Zr, 0.3 g Te. Iodine of 2 mg/cm$^3$ is employed as the transport agent. As illustrated in \rfig{fig.sem}a, source materials, iodine and the growth substrates are sealed in a quartz ampoule, which has been flushed and pumped to 5 Pa. In particular, zirconium shots and iodine particles are placed in one side of the ampoule, while tellurium powder and silicon substrates are placed in the other side. The iodine vapor transport growth of ZrTe$_5$ can be qualitatively described by the following reactions:
\begin{align*}
\ce{Zr + 2I_2 &-> ZrI4}\\
\ce{ZrI4 + 5Te &-> ZrTe5 + 2I2}
\end{align*}
. Due to the low vapor pressure of Zr, the first reaction is set in the highest temperature zone, so sufficient ZrI$_4$ can be fed to the second reaction, which is held in the lowest temperature zone. The tellurium powder is placed in the third temperature zone in between the other two zones. The optimal temperatures for three zones were found to be 540 $^\circ$C, 480 $^\circ$C and 450 $^\circ$C, respectively. The growth usually takes place for 150 min to grow nanostructures(or 90 min for bulk), and then the ampoule is allowed to cool naturally.

Nanoribbons were found on the silicon substrate and certain segments of the ampoule wall, as shown by the SEM image in \rfig{fig.sem}. These nanostructures can be very long, e.g., 200 microns. Later we will show via scanning transmission electron microscopy(STEM) and Raman spectroscopy that these long nanostructures are ZrTe$_5$. A substantial advantage of our method is the fast growth rate. Depending on the growth conditions, such as the ratio and the temperature of the source materials, macroscopic crystals can also been grown in similar growth time on the ampoule wall. The size of crystals can be as large as 10 mm by 0.2 mm by 0.05 mm, rivalling reported results with several weeks of growth time\cite{Li2014,Sodeck1979,Levy1983}. A rough estimation yields a growth rate being at least 40 times faster.

Interestingly, these ZrTe$_5$ nanostructures did not directly grow on the silicon substrate, but on a mattress of materials, as seen in \rfig{fig.sem}b. Energy-dispersive X-ray spectroscopy(EDX) measurements show that the mattress mainly consists of Zr and Te with a ratio of $\sim$1:3, indicating ZrTe$_3$(see the supporting information). Some tellurium crystals are also found. Apparently, ZrTe$_3$ and Te crystals grow first. They may provide a suitable substrate for subsequent growth of ZrTe$_5$. A close look at the edge of the mattress, as seen in \rfig{fig.sem}d and e, reveals recess etching into the silicon substrate. Iodine is known to react with silicon and form SiI$_4$, which is in a gas phase at our growth temperature. We have found that silicon plays an important role in the reaction. When the silicon substrate is replaced with mica, neither ZrTe$_3$ nor ZrTe$_5$ formed at the same growth condition. On the other hand, in presence of both silicon and mica substrates, growth can occur on the mica substrate, though much less effectively. How silicon affects the growth processes is not clear currently and deserves further study. In the following, we focus on the quality of the grown nanoribbons.

\begin{figure}[htbp]
\includegraphics[width=1\columnwidth]{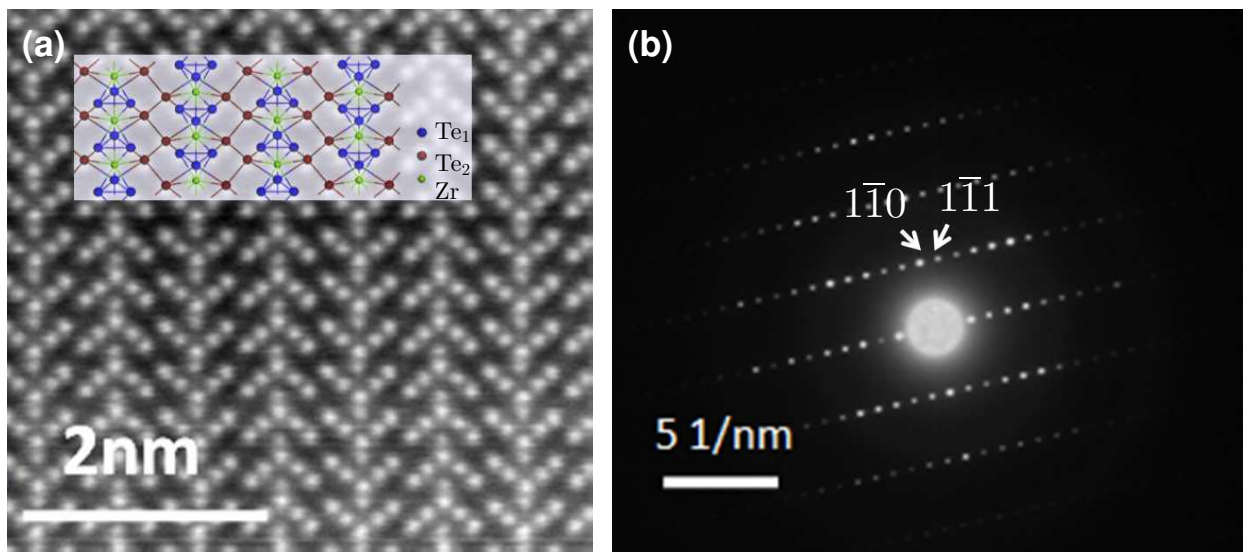}
\caption{Aberration corrected TEM images of ZrTe$_5$ nanoribbon. The nanoribbon is about 50 nm thick. (a)HAADF image. Inset, atomic structure model of ZrTe$_5$ in the (110) plane. (b)Electron diffraction pattern along the [110] direction.}
\label{fig.tem}
\end{figure}

EDX was carried out for nanoribbons, which indicates that they are made of Zr and Te. The atomic ratio is about 1:5(within 0.5\% error). High-angle annular dark-field(HAADF) images were taken with the aberration corrected transmission electron microscope(JEOL ARM200F). \rfig{fig.tem} is such an image of a typical nanoribbon. The most prominent feature is the vertical arrow-like atomic chains. The image matches well with the expected atomic arrangement of the (110) plane of ZrTe$_5$ depicted in the inset of \rfig{fig.tem}a, proving that these nanostructures are ZrTe$_5$. The electron diffraction pattern looking down the [110] direction in \rfig{fig.tem}b also agrees well with ZrTe$_5$. Furthermore, we find that the lattice constants, $a=0.40$ nm, $b=1.45$ nm and $c=1.34$ nm, are very closed to the expected value for ZrTe$_5$, as well\cite{Li2014,Niu2017,Yu2016,Weng2014}. From these images, it can be seen that the grown samples are of high crystalline quality.

All data show that nanoribbons are along the $a$-axis, and the shortest dimension is along the $b$-axis. This growth mode is in fact expected considering the crystal structure. ZrTe$_5$ is a layered material coupled by Van der Waals interactions, which favors 2D growth in principle. In each layer, there is a strong structural anisotropy, which leads to the dimension along the $a$-axis being significantly longer than the other dimension (along the $c$-axis).

\begin{figure}[htbp]
\includegraphics[width=1\columnwidth]{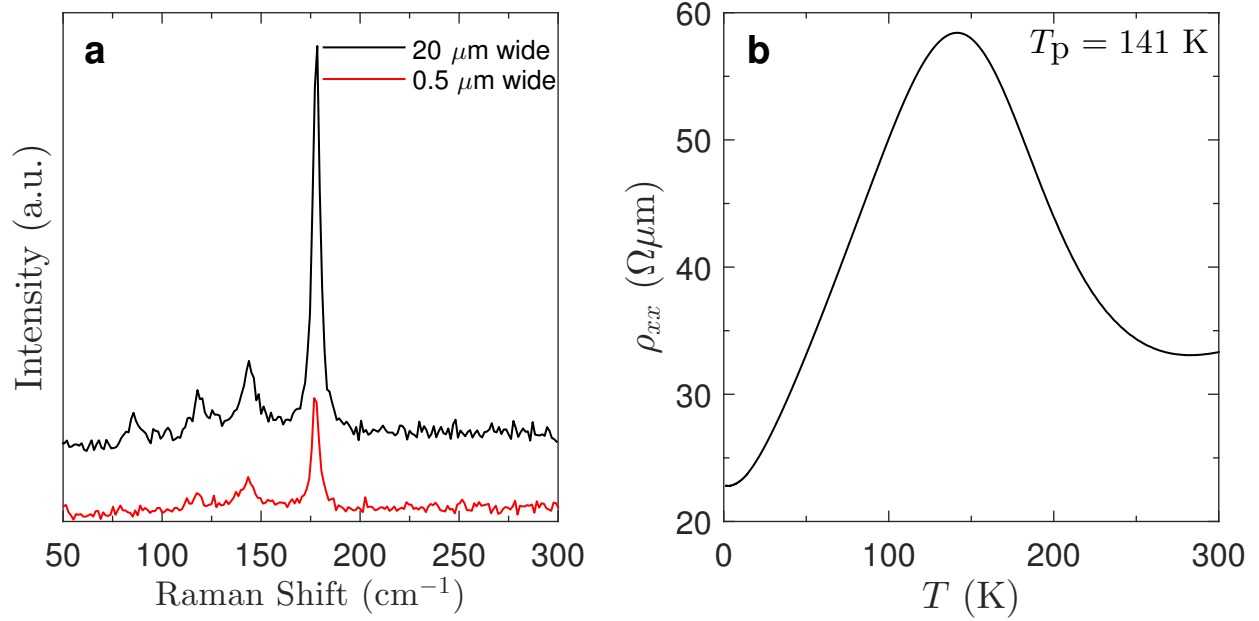}
\caption{(a)Raman spectra of ZrTe$_5$ nanoribbons. Four peaks at 115, 119, 145 and 179 cm$^{-1}$ are well resolved in the 20 $\mu$m wide ribbon, while the signal is weaker in the 0.5 $\mu$m wide ribbon.(b)Temperature dependence of resistivity for a 10 $\mu$m by 400 nm by 90 nm ZrTe$_5$ nanoribbon. A broad maximum appears at $T_\text{p}=141$ K.}
\label{fig.rtandraman}
\end{figure}

Raman spectroscopic measurements have been performed. The Raman spectra for narrow and wide nanoribbons are similar, except that narrow ones have a weaker signal due to a smaller volume to interact with light. Four characteristic peaks, centered at 115, 119, 145 and 179 cm$^{-1}$, can be readily identified as vibration modes of Te atoms\cite{Taguchi1983}. The peak positions agree with the bulk as shown in \rfig{fig.rtandraman}a.

The temperature dependence of resistivity of nanoribbons, shown in \rfig{fig.rtandraman}b, exhibits a maximum at 141 K. Such a nonmonotonic behavior is characteristic for ZrTe$_5$\cite{Okada1982,DiSalvo1981,Tritt1999,Manzoni2015,Chi2017}. Although different explanations have been proposed, its origin is still under debate. Recently, we have shown in exfoliated ZrTe$_5$ flakes that the resistivity maximum likely results from competition between a Dirac semimetal band and a semiconductor band\cite{Niu2017}.

\begin{figure}[htbp]
\includegraphics[width=1\columnwidth]{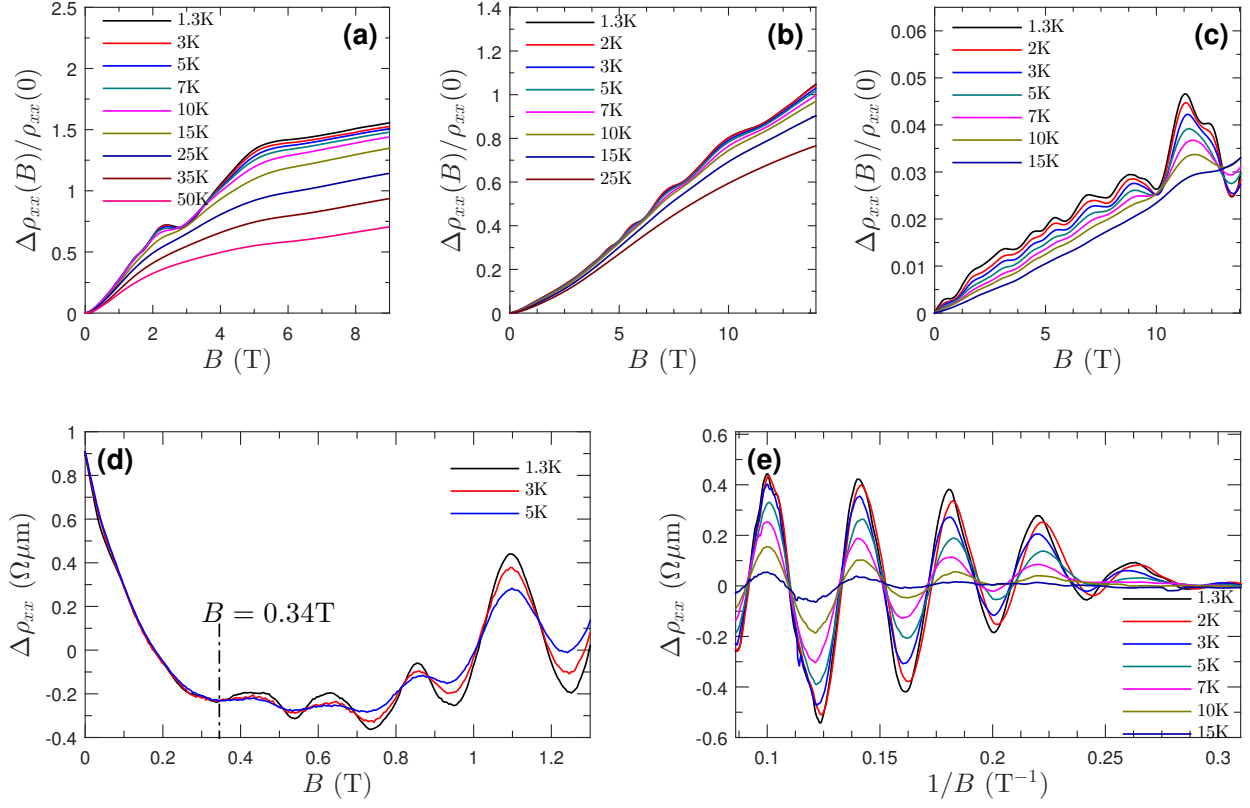}
\caption{Magnetoresistance for a ZrTe$_5$ nanoribbon. The current is along the $a$-axis. (a)$\mathbf{B}\parallel \mathbf{b}$. (b)$\mathbf{B}\parallel \mathbf{c}$. (c)$\mathbf{B}\parallel \mathbf{a}$.(d)Shubnikov-de Haas oscillations with a smooth background subtracted along the $b$-axis. (e)Shubnikov-de Haas oscillations with a smooth background subtracted along the $c$-axis.}
\label{fig.mr}
\end{figure}

The quality of the nanoribbon is manifested in electrical transport. We have carried out magnetoresistance(MR) measurements under magnetic fields in different directions, as depicted in \rfig{fig.mr}. The current is always along the nanoribbon, the $a$-axis. When the field is perpendicular to the current, MR is significant, while it is very small when the field is parallel to the current. In all three field directions, MR displays marked oscillations, the so-called Shubnikov-de Haas oscillations(SdHOs). Surprisingly, SdHOs start to appear in a field $B_\text{min}$ as low as 0.34 T as shown in \rfig{fig.mr}d, which suggests a high electrical mobility. Note that SdHOs stem from formation of Landau levels, which requires a condition of $\omega_\text{c}\tau>1$. Here $\omega_\text{c}$ is the cyclotron frequency, and $\tau$ the mean free path. It is easy to show that the condition is equivalent to $\omega_\text{c}\tau=\mu B>1$, where $\mu$ is the mobility. Plugging $B_\text{min}=0.34$ T, we have $\mu \sim 3 \times 10^4$ cm$^2$ V$^{-1}$ s$^{-1}$. The high mobility will surely be very helpful in studying the transport properties of this topological material. Therefore, our work provides a promising growth method for high mobility nanostructured ZrTe$_5$.

\begin{figure}[htbp]
\includegraphics[width=1\columnwidth]{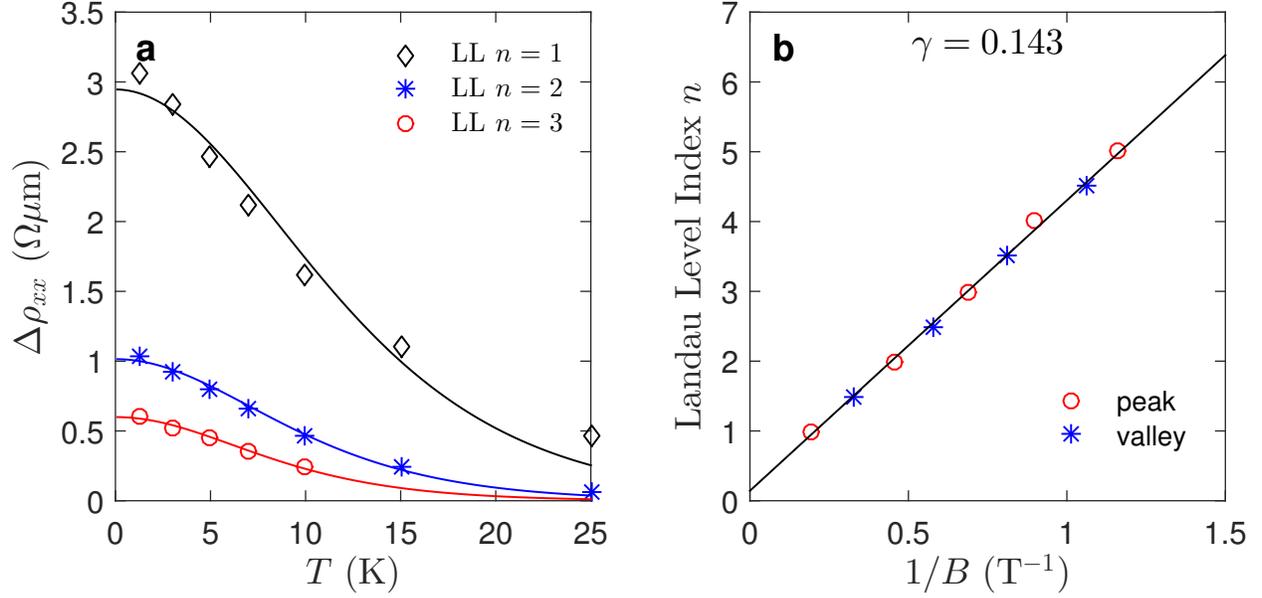}
\caption{Analysis of SdH oscillations. (a)Temperature dependence of the oscillation amplitudes. Symbols are experimental data, while lines are best fits to \req{eq.LK}. Black, blue and red are for Landau levels $n=1$, $n=2$ and $n=3$, respectively. (b)Landau plot of the oscillations. The black line is a linear fit.}
\label{fig.landau}
\end{figure}

In \rfig{fig.mr}a, it can be seen that the quantum limit is reached at a relatively small field, 5.5 T, indicating a low carrier density. The low doping level also suggests high quality of the sample, corroborating with other measurements. After subtracting a smooth background, the oscillations are clearly resolved, shown in \rfig{fig.mr}e. By picking the field positions of the maxima and minima of the oscillations and plotting the inverse position against the Landau level index $n$, we obtain the Landau plot, shown in \rfig{fig.landau}a. Here, the maxima are assigned  integer indices, while the minima are assigned half integer indices. Under this convention, an intercept of $\gamma\sim0.143$ on the $y$-axis is estimated from a linear fit. The intercept of the Landau plot has been widely used to determine the non-trivial Berry phase of $\pi$ for Dirac systems.\cite{Murakawa2013} For a 2D Dirac system, like graphene, $\gamma=0$, while for 3D materials with a Dirac cone, the intercept is expected to be $\pm 1/8$\cite{Liu2016,Murakawa2013,Yuan2015,Yu2016}. It is positive for holes and negative for electrons.\cite{Qu2010} Therefore, our data are in excellent agreement with a Dirac system with dominant hole carriers, confirming previous studies.\cite{Zheng2016,Yu2016,Chen2017}

The slope of the linear dependence gives the oscillation frequency $B_\text{f}$. For three magnetic field orientations, $B_\text{f}=$30.33, 5.14 and 25.67 T, respectively. According to the Lifshitz-Onsager relation, $B_\text{f}=\hbar S_\text{e}/2\pi e$, where $\hbar$ and $e$ are the reduced Plank constant and the elementary charge, respectively, and $S_\text{e}$ is the extremal cross-sectional area of the Fermi surface in a plane normal to the magnetic field. Adopting an ellipsoidal Fermi surface, as suggested in earlier studies,\cite{Kamm1985,Zheng2016,Zheng2016a} the Fermi wave vectors along three principle axes, can be estimated as $k_\text{a}=0.023$ nm$^{-1}$, $k_\text{b}=0.136$ nm$^{-1}$ and $k_\text{c}=0.0272$ nm$^{-1}$. Furthermore, the damping of the oscillation amplitude $A$ with temperature can provide an estimation of the band velocity $v_0$ of the massless Dirac fermion in the system, based on the Lifshitz-Kosevich relation
\begin{equation}
\frac{A(T)}{A(0)}=\lambda/\sinh(\lambda), \lambda=\frac{2\pi^2k_\texttt{B}Tk_\text{F}}{ev_0B}
\label{eq.LK}
\end{equation}
, where $k_\texttt{B}$ is the Boltzmann constant. In \rfig{fig.landau}b, $A(T)/A(0)$ for $\mathbf{B}\parallel \mathbf{b}$ are plotted against temperature for several Landau levels, $n=$1, 2 and 3. By fitting the plots to \req{eq.LK}, we obtain the velocity along the $b$-axis, $v_0^b\approx4.7\times10^5$ ms$^{-1}$, in agreement with values reported by others.\cite{Yuan2015,Zheng2016}

In conclusion, we report growth of ZrTe$_5$ nanoribbons by a silicon-assisted chemical vapor transport technique. Compared with the previous growth method for bulk crystals, our technique has the advantage of significantly shorter growth time. The grown nanoribbons are of high crystalline quality and display low unintentional doping and high mobility, suitable for study of topological properties near the charge neutrality point and beyond the quantum limit. Quantum transport experiments indicate that the ZrTe$_5$ nanoribbon is a Dirac semimetal with a nontrivial Berry phase $\pi$ in the $ac$ plane. Our work is the first experiment on growth of ZrTe$_5$ nanostructures and provides a good starting point for studies of nanostructured ZrTe$_5$.

\begin{acknowledgement}
This work was supported by the National Key Research and Development Program of China (No. 2016YFA0300600, No. 2013CBA01603, No. 2016YFA0300802, No. 2013CB932904 and No. 2016YFA0202500) and NSFC (Project No. 11574005, No. 11222436, and No. 11234001).
\end{acknowledgement}

%\bibliography{ZrTe5.growth}
\providecommand{\latin}[1]{#1}
\makeatletter
\providecommand{\doi}
  {\begingroup\let\do\@makeother\dospecials
  \catcode`\{=1 \catcode`\}=2 \doi@aux}
\providecommand{\doi@aux}[1]{\endgroup\texttt{#1}}
\makeatother
\providecommand*\mcitethebibliography{\thebibliography}
\csname @ifundefined\endcsname{endmcitethebibliography}
  {\let\endmcitethebibliography\endthebibliography}{}

\newpage

\newpage

\newpage

\newpage

\clearpage

\newpage

%%%%%%%%%% Merge with supplemental materials %%%%%%%%%%
\pagebreak

\begin{center}
\textbf{\large Supplementary materials: Facile and fast growth of high mobility nanoribbons of ZrTe$_5$}
\end{center}
%%%%%%%%%% Merge with supplemental materials %%%%%%%%%%
%%%%%%%%%% Prefix a "S" to all equations, figures, tables and reset the counter %%%%%%%%%%
\setcounter{equation}{0}
\setcounter{figure}{0}
\setcounter{table}{0}
\setcounter{page}{1}
\makeatletter
\renewcommand{\theequation}{S\arabic{equation}}
\renewcommand{\thefigure}{S\arabic{figure}}
\renewcommand{\bibnumfmt}[1]{[S#1]}
\renewcommand{\citenumfont}[1]{S#1}
%%%%%%%%%% Prefix a "S" to all equations, figures, tables and reset the counter %%%%%%%%%%

\begin{figure}[htbp]
\includegraphics[width=1\columnwidth]{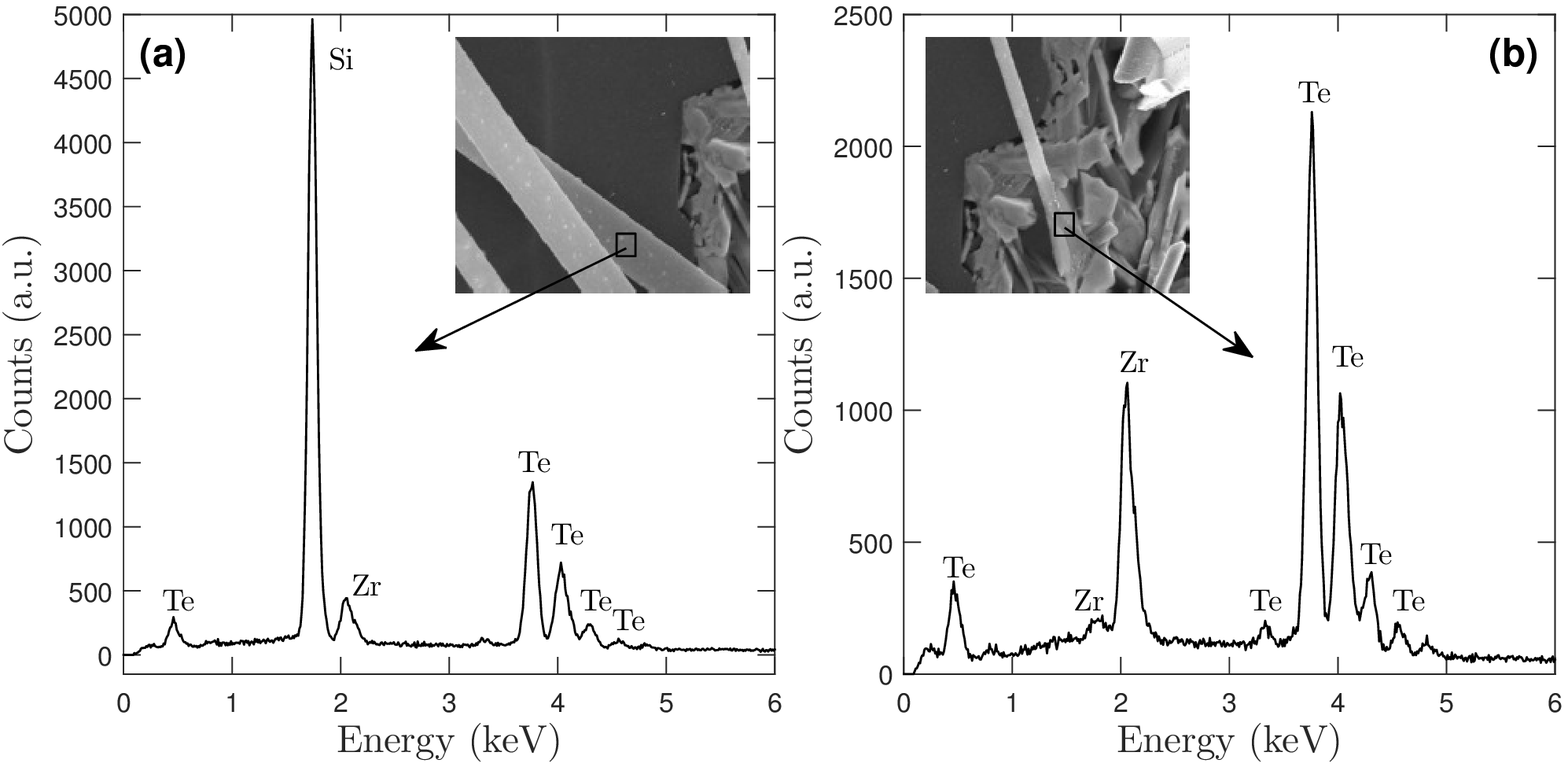}
\caption{Energy-dispersive X-ray spectroscopy of materials grown on silicon substrates. (a)A spectrum taken on nanoribbon, indicating ZrTe$_5$. (b)A spectrum taken at the root of nanoribbon, indicating ZrTe$_3$. The insets are SEM images. The black squares mark the spots where the spectra were taken.}
\label{sfig.edx}
\end{figure}

\newpage

\begin{figure}[htbp]
\includegraphics[width=1\columnwidth]{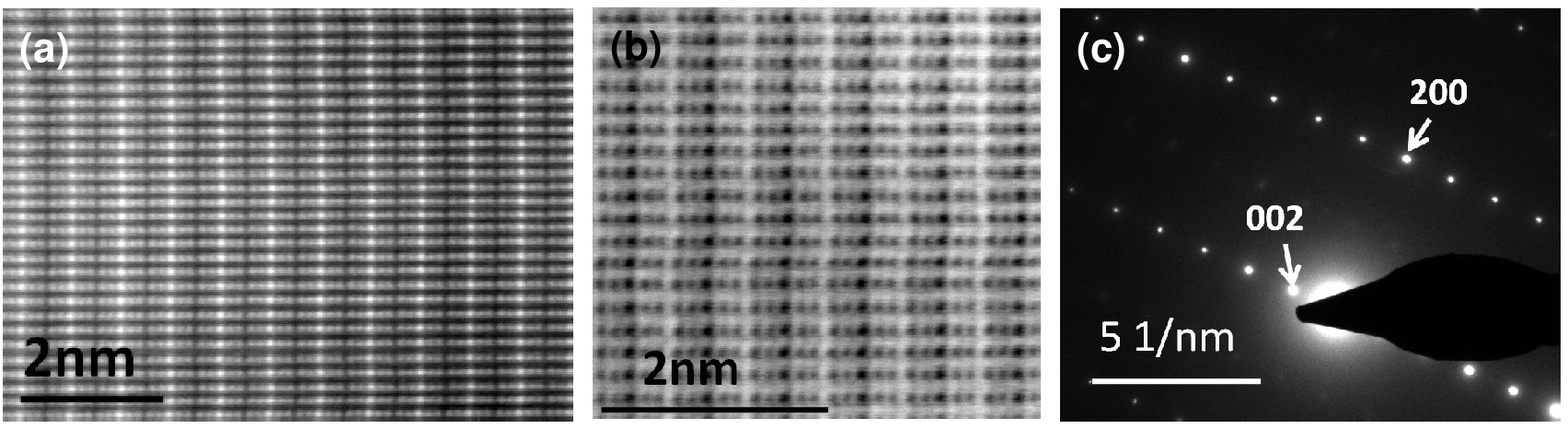}
\caption{Aberration corrected TEM images of ZrTe$_5$ nanoribbon along the [010] direction. (a)High-angle annular dark-field image. (b)High-angle annular light-field image. (c)Electron diffraction pattern.}
\label{sfig.stem010}
\end{figure}

\newpage

\begin{figure}[htbp]
\includegraphics[width=1\columnwidth]{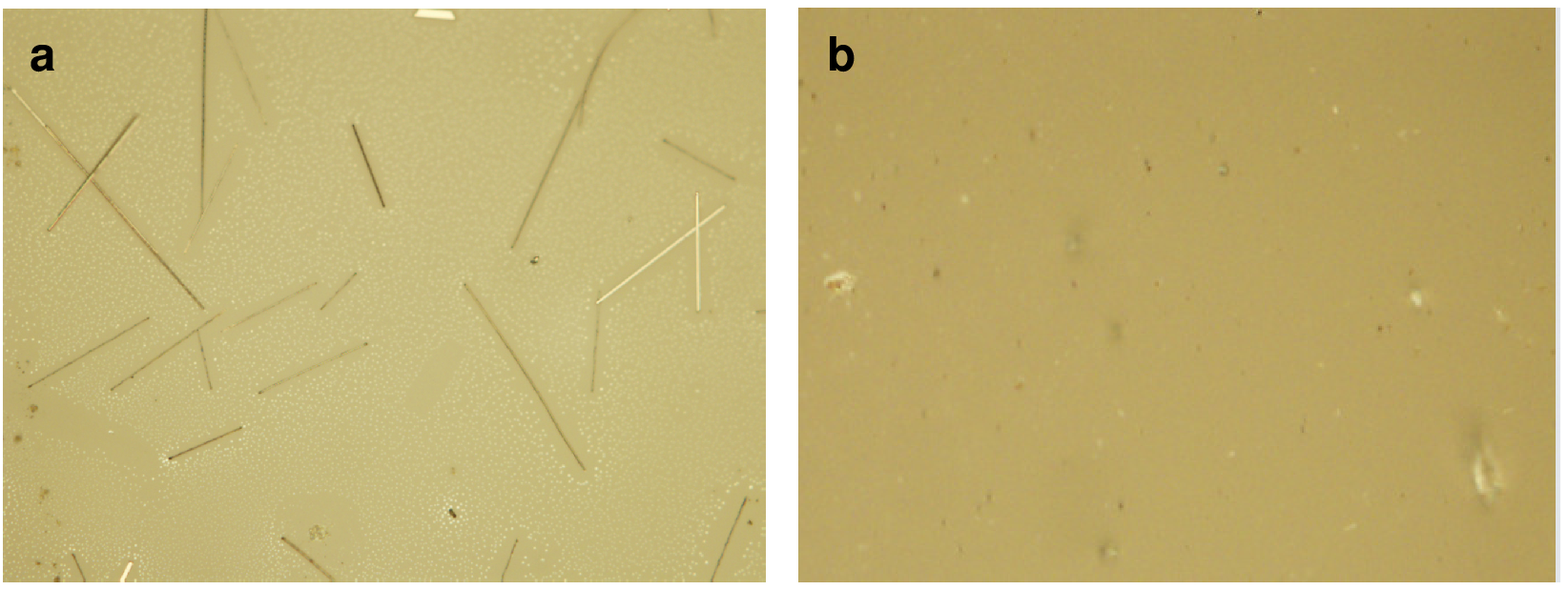}
\caption{(a)In presence of both silicon and mica substrates, ZrTe$_5$ can grow on the mica substrate. (b)Without the existence of silicon, ZrTe$_5$ cannot grow on mica substrate}
\label{sfig.micaall}
\end{figure}

\newpage

\begin{figure}[htbp]
\includegraphics[width=0.6\columnwidth]{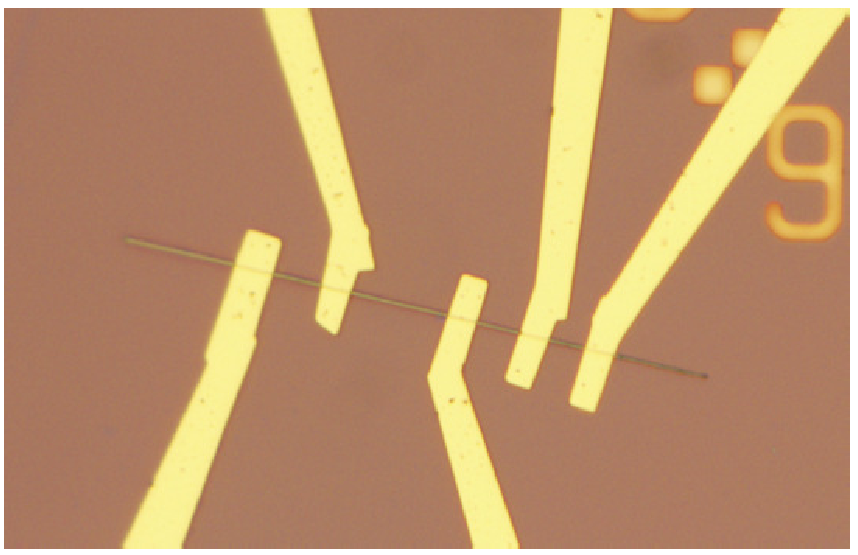}
\caption{Optical micrograph of a typical four-point measurement device of a ZrTe$_5$ nanoribbon.}
\label{sfig.device}
\end{figure}

\newpage

\begin{figure}[htbp]
\includegraphics[width=1\columnwidth]{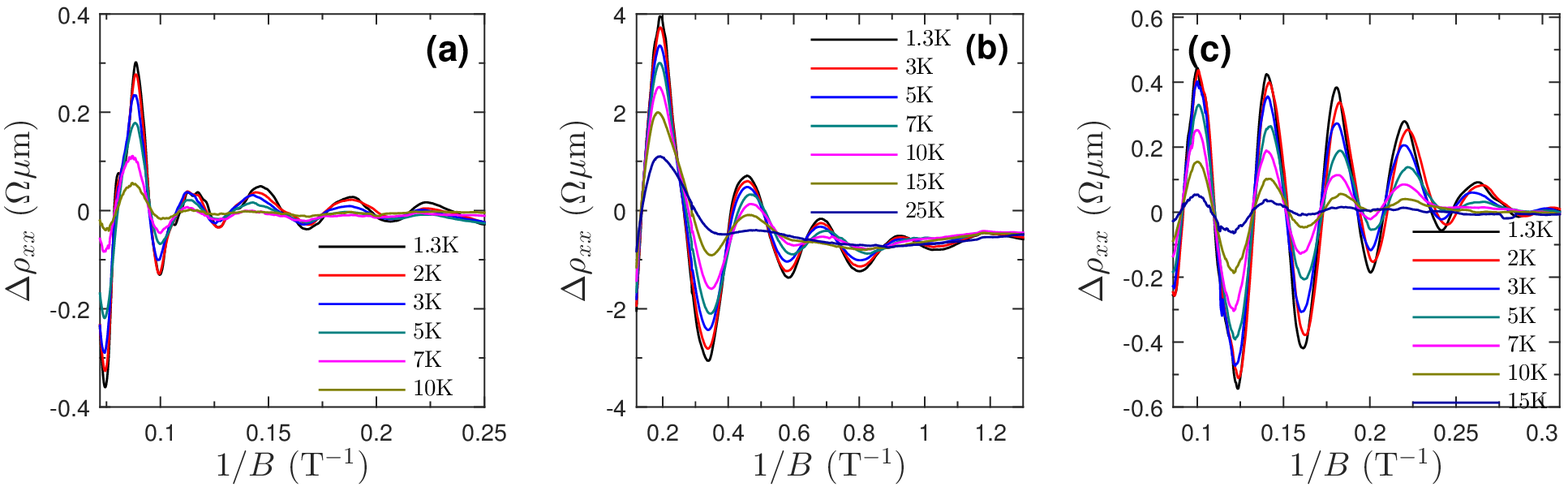}
\caption{SdH oscillations with the magnetic field along three principal directions at various temperatures. A smooth background has been subtracted. (a)$\mathbf{B}\parallel \mathbf{a}$. (b)$\mathbf{B}\parallel \mathbf{b}$. (c)$\mathbf{B}\parallel \mathbf{c}$.}
\label{sfig.3dsdh}
\end{figure}

\newpage

\begin{figure}[htbp]
\includegraphics[width=0.6\columnwidth]{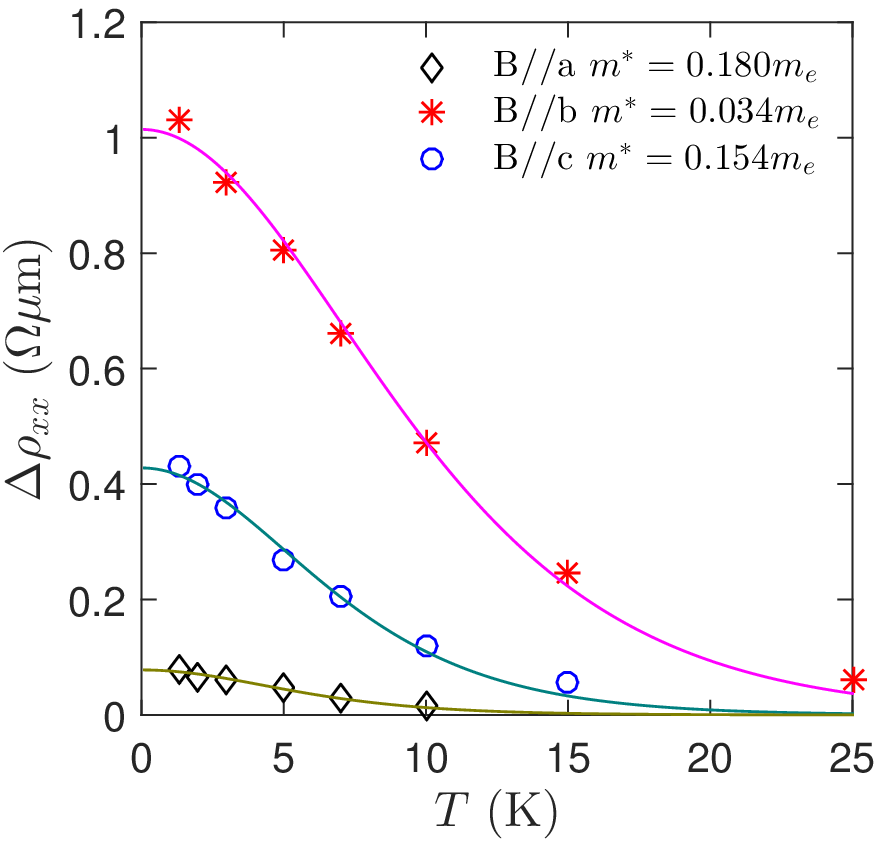}
\caption{Temperature dependence of the SdH oscillation amplitude along three axes. Open diamond, $\mathbf{B}\parallel \mathbf{a}$. Asterisk, $\mathbf{B}\parallel \mathbf{b}$. Open circle, $\mathbf{B}\parallel \mathbf{c}$. Solid lines are fits to the Lifshitz-Kosevich relation(Eq.~1 in the main text), from which the cyclotron mass are estimated as $m^*_a=0.180m_\text{e}$, $m^*_b=0.034m_\text{e}$ and $m^*_c=0.154m_\text{e}$, respectively.}
\label{sfig.lk}
\end{figure}

\end{document}